\documentclass[preprint]{revtex4}

\def\al{\alpha}
\def\be{\beta}

\def\de{\delta}

\def\th{\theta}

\def\ch{\chi}
\def\ps{\psi}

\def\La{\Lambda}

\def\fr#1#2{{{#1} \over {#2}}}

\def\frac#1#2{{\textstyle{{#1}\over {#2}}}}

\def\vev#1{\langle {#1}\rangle}

\def\lsim{\mathrel{\rlap{\lower4pt\hbox{\hskip1pt$\sim$}}
   \raise1pt\hbox{$<$}}}
\def\gsim{\mathrel{\rlap{\lower4pt\hbox{\hskip1pt$\sim$}}
   \raise1pt\hbox{$>$}}}
\def\sqr#1#2{{\vcenter{\vbox{\hrule height.#2pt
        \hbox{\vrule width.#2pt height#1pt \kern#1pt
        \vrule width.#2pt}
        \hrule height.#2pt}}}}

\newcommand{\beq}{\begin{equation}}
\newcommand{\eeq}{\end{equation}}
\newcommand{\bea}{\begin{eqnarray}}
\newcommand{\eea}{\end{eqnarray}}
\newcommand{\bit}{\begin{itemize}}
\newcommand{\eit}{\end{itemize}}
\newcommand{\rf}[1]{(\ref{#1})}

\begin{document}

\title{Gravity with background fields and diffeomorphism breaking}

\author{Robert Bluhm}

\affiliation{
Physics Department, Colby College,
Waterville, ME 04901 
}

%\date{January 2015}

\begin{abstract}
Effective gravitational field theories with background fields
break local Lorentz symmetry and diffeomorphism invariance.
Examples include Chern-Simons gravity, massive gravity,
and the Standard-Model Extension (SME).
The physical properties and behavior of these theories 
depend greatly on whether the spacetime symmetry
breaking is explicit or spontaneous.
With explicit breaking, the background fields are fixed and nondynamical,
and the resulting theories are fundamentally different 
from Einstein's General Relativity (GR).
However, when the symmetry breaking is spontaneous,
the background fields are dynamical in origin,
and many of the usual features of Einstein's GR still apply.
\end{abstract}

%\pacs{11.30.Cp, 11.10.Ef, 04.40.Nr}

\maketitle

\section{Introduction}

Ideas originating out of quantum gravity and string theory suggest that
Lorentz symmetry and diffeomorphism invariance might not hold as exact 
symmetries in nature across all energy scales.
At the same time, searches for alternative theories that could explain
dark energy and dark matter consider the possibility of modifications
to Einstein's General Relativity (GR).

Together these types of ideas have led to a number of effective gravitational
field theories being proposed in which 
Lorentz symmetry and diffeomorphism invariance are broken.
Typically, it is the presence of fixed background fields introduced at the
level of effective field theory that indicates breaking of spacetime symmetry.
These backgrounds are associated with either explicit breaking or
spontaneous breaking of diffeomorphism invariance.

This brief review looks at both of these types of diffeomorphism breaking
and compares the properties and features of the resulting theories to Einstein's GR.

\section{Einstein's GR and background fields}

Consider Einstein's GR with an action containing an Einstein-Hilbert term and
minimally coupled matter fields,
\beq
S = \int d^4 x \, \sqrt{-g} \left( \fr 1 {16 \pi G} R + {\cal L}_{\rm M} (g_{\mu\nu}, f^\ps) \right) .
\eeq
Here, $f^\ps$ denotes generic matter fields with spacetime labels 
written collectively as $\ps$.
The Einstein equations obtained by varying with respect to the metric are
$G^{\mu\nu} = 8 \pi G \, T^{\mu\nu}_{\rm M}$.

Standard properties of GR include the following:
The theory is invariant under diffeomorphisms involving a vector $\xi^\mu$,
and physical solutions for the metric form equivalence classes
related by these transformations.
The contracted Bianchi identity, $D_\mu G^{\mu\nu} = 0$,
reduces the ten Einstein equations to six dynamically independent equations
for the metric tensor.
Of the six possible dynamical metric modes,
four are eliminated as diffeomorphism gauge degrees of freedom.
The equations $D_\mu T^{\mu\nu}_{\rm M} = 0$ that follow from
the contracted Bianchi identity and Einstein's equations
are satisfied by the dynamical degrees of freedom
associated with the matter fields $f^\ps$, not the metric.
The result is that in GR the geometry described by the metric
is influenced by the dynamics of the matter fields and vice versa.

When fixed background fields are added to GR
diffeomorphism invariance is broken,
and the linkage between spacetime geometry and
the dynamics of the nongravitational fields in the theory is disturbed.
Background fields must be treated differently from conventional matter fields.
Previous studies refer to them as ``absolute objects,''  
which cannot have backreactions.\cite{A67}

When background fields are included in a gravitational theory at the
level of effective field theory,
there is, however, an important distinction that must be made
between whether diffeomorphisms are broken explicitly 
versus spontaneously.\cite{akgrav04,rb15a}

To examine this distinction and for concreteness, 
consider the modified action,
\beq
S = \int d^4 x \, \sqrt{-g} \left( \fr 1 {16 \pi G} R + {\cal L}_{\rm M} (g_{\mu\nu}, f^\ps) 
+ {\cal L}_{\rm LV} (g_{\mu\nu}, f^\ps, \bar k_\ch)\right) ,
\label{S}
\eeq
where an additional Lorentz- and diffeomorphism-violating term,
$ {\cal L}_{\rm LV}$, has been added.
The extra term depends on a fixed background field written here as $\bar k_\ch$, 
with spacetime indices denoted collectively as $\ch$.
The Einstein equations,$G^{\mu\nu} = 8 \pi G \, (T^{\mu\nu}_{\rm M} + T^{\mu\nu}_{\rm LV})$,
acquire an extra contribution as well from the additional term.

Under diffeomorphisms,
the metric and conventional matter fields, $f^\ps$, transform
with changes given by their Lie derivatives.
For example, 
\beq
g_{\mu\nu} \rightarrow g_{\mu\nu} + {\cal L}_{\xi} g_{\mu\nu}
= g_{\mu\nu} + D_\mu \xi_\nu + D_\nu \xi_\mu .
\label{metricdiff}
\eeq
However,
the background $\bar k_\ch$ remains fixed under diffeomorphisms
and does not transform.
It is because of this behavior that the action $S$ is not invariant under diffeomorphisms,
and $(\de S)_{\rm diffs} \ne 0$.

At the same time, to maintain observer invariance, the action $S$ must
be invariant under general coordinate transformations.
These includes coordinate transformations given as 
$x^\mu \rightarrow x^{\mu^\prime} (x)$.
By choosing $x^{\mu^\prime}(x)$ as an infinitesimal 
coordinate transformation to $x^\mu - \xi^\mu$, 
using an opposite sign for $\xi^\mu$,
and by performing Taylor expansions in the Lagrangian density,
a set of general coordinate transformations that mathematically 
have the same form as the diffeomorphisms can be found.
For example, under these transformations the metric again transforms
as in \rf{metricdiff}.
However, here the difference is that under these observer transformations
the background does transform, and it obeys
$\bar k_\ch \rightarrow \bar k_\ch + {\cal L}_{\xi} \bar k_\ch$.
Therefore, in this case, the action $S$ is invariant under these
observer transformations,
and $(\de S)_{\rm GCTs} = 0$.

The fact that $S$ is not invariant under diffeomorphisms,
while it must be invariant under the observer general coordinate transformations
gives rise to a potential inconsistency.\cite{akgrav04,rb15a}  
The next sections examine this potential conflict for the cases of
when the diffeomorphism breaking is explicit versus when it occurs as
a result of spontaneous symmetry breaking.

\section{Explicit diffeomorphism breaking}

When diffeomorpsim breaking is explicit it is due to the fact that
the background field does not arise dynamically and does not have equations of motion.
Instead, it is included as an absolute object,
which is not able to have backreactions.

As a result, mathematically, variations in the action with respect to the 
background $\bar k_\ch$ in \rf{S} need not vanish,
and
\beq
 \int d^4x 
\sqrt{-g} \fr {\de {\cal L}_{\rm LV}} {\de \bar k_\ch} \de \bar k_\ch
\ne 0 
\quad\quad ({\rm explicit \,\, breaking}) .
\label{breaking}
\eeq

At the same time, the theory must be invariant under the observer general 
coordinate transformations described above.
When these are performed,
using integration by parts for the terms involving 
${\cal L}_\xi g_{\mu\nu} = D_\mu \xi_\nu + D_\nu \xi_\mu$,
and when the dynamical equations of motion for the matter fields $f^\ps$ are imposed,
the result is 
\beq
 \int d^4x \sqrt{-g}  \, 
 \left[ D_\mu \left( T^{\mu\nu}_{\rm M} + T^{\mu\nu}_{\rm LV} \right) \xi_\nu
 -  \fr {\de  {\cal L}_{\rm LV}} {\de \bar k_\ch} {\cal L}_\xi \bar k_\ch
 \right] = 0.
 \label{nobreaking}
 \eeq

Because of these results,
a potential conflict arises when the symmetry breaking is explicit.
If the divergence of Einstein's equations is taken and the contracted
Bianchi identity is used,
then $D_\mu \left( T^{\mu\nu}_{\rm M} + T^{\mu\nu}_{\rm LV} \right) = 0$
holds on shell,
and the first term in \rf{nobreaking} vanishes.
This leaves the requirement that the integral in \rf{breaking} must vanish on shell as well,
which would appear to contradict the statement that diffeomorphism invariance is broken
and the background is nondynamical.
 
There appear to be only three possibilities in the case
with explicit breaking.\cite{rb15a}
One is that the theory is inconsistent and
the fixed background $ \bar k_\ch$ must therefore vanish.
The second is that the integrands in \rf{breaking} and \rf{nobreaking}
vanish on shell despite the fact that $\bar k_\ch$ is not dynamical.
The third is that the integrand in \rf{nobreaking}
equals a total derivative,
allowing the integral to vanish 
off shell even when the
integral in \rf{breaking} does not vanish.

Which of these three possibilities comes into play depends in large part on
the tensor nature of the background $\bar k_\ch$.
The most restrictive cases occur when the background is a fixed scalar function $\bar k$.
With a background scalar,
the Lie derivative ${\cal L}_\xi \bar k = - \xi^\mu \partial_\mu \bar k$,
and an overall factor of $\xi^\mu$ can be pulled out of the integrand in \rf{nobreaking}.
Since the integral must then vanish for all $\xi^\mu$,
the result is that 
\beq
D_\mu \left( T^{\mu\nu}_{\rm M} + T^{\mu\nu}_{\rm LV} \right) 
 = - \fr {\de  {\cal L}_{\rm LV}}  {\de \bar k} \partial_\mu \bar k .
 \label{scalar}
 \eeq 
When the Einstein equations are put on shell,
the left-hand side of this equation must vanish.
Assuming that $\bar k$ is not a constant then it must be that the 
variation $\fr {\de  {\cal L}_{\rm LV}}  {\de \bar k}$ must vanish
or else the theory is inconsistent.

An example of an inconsistent model involving a scalar field is when
an explicit time-dependent cosmological constant $\La(t)$ is added to GR.
In this case, with the scalar $\bar k = \La(t)$,
the variation $\fr {\de  {\cal L}_{\rm LV}}  {\de \La} \ne 0$,
and the only option is that $\partial_\mu \La(t) = 0$.
However, with explicit time dependence in $\La(t)$ this does not hold,
and the theory is therefore inconsistent.

A second example with a scalar background is Chern-Simons gravity,\cite{rjsp}
where 
\beq
\sqrt{-g} {\cal L}_{\rm LV} = \fr 1 {64 \pi G} \, \th \, {^\ast}R R .
\eeq
Here, ${^\ast}R R$ is the gravitational Pontryagin density and $\th$
is a fixed scalar background.
With $\bar k = \th$ the variation $\fr {\de  {\cal L}_{\rm LV}}  {\de \bar k}$
is found to be directly proportional to the Pontryagin density ${^\ast}R R$.
As a result, the inconsistency can be evaded by restricting
the solutions for the metric to spacetimes that have a vanishing
Pontryagin density.
In this example, $\fr {\de  {\cal L}_{\rm LV}}  {\de \bar k} = 0$ holds on shell
despite the fact that $\bar k = \th$ is not dynamical.  

With scalar backgrounds it is also possible to construct models that
evade the inconsistency by imposing constraints on other fields besides the metric.
An example of this technique is used to show that a tensor-vector theory with explicit
diffeomorphism breaking can give rise to Einstein-Maxwell solutions,
where the consistency condition effectively imposes
a gauge-fixing condition on the vector field.\cite{rb15b}

On the other hand, if the background $\bar k_\ch$ is a tensor field,
there is an additional way to evade the potential inconsistency.
With a tensor background,
the Lie derivative acting on it includes contributions with derivatives acting on $\xi^\mu$.
Using integration by parts on these contributions,
it can be shown in general that the integrand appearing in \rf{nobreaking}
becomes a total derivative.  
This permits evasion of the potential inconsistency because the integral in \rf{nobreaking}
can then vanish even when the variations in \rf{breaking} do not.

An example of a theory with a tensor background is massive gravity.\cite{massive} 
In these theories,
a background field that is a symmetric two-tensor,
written here as $\bar k_{\mu\nu}$,
is used to create mass terms for the metric.
The explicit symmetry-breaking term, ${\cal L}_{\rm LV}$,
in this case is only a function of the metric and the background $\bar k_{\mu\nu}$.
With terms of this form, the energy-momentum tensor can be shown to obey 
an off-shell relation of the form
\beq
(D_\mu T_{\rm LV}^{\mu\nu}) \xi_\nu 
+ \fr {\de  {\cal L}} {\de \bar k_{\mu\nu}} {\cal L}_\xi \bar k_{\mu\nu}
= D_\mu \left(2 \fr {\de {\cal L}_{\rm LV}} {\de g^{\al\be}} g^{\mu\al} \xi^\be \right)  .
\label{DTU4}
\eeq
Assuming the energy-momentum tensor for the matter fields separately obeys 
$D_\mu T_{\rm M}^{\mu\nu} = 0$ on shell,
it then follows as a result of \rf{DTU4} that the
integrand appearing in \rf{nobreaking} is a total derivative.
As a result, the integral \rf{nobreaking} vanishes even though
\rf{breaking} need not vanish off shell.

Although these examples with nondynamical backgrounds 
are able to evade the potential inconsistency associated with explicit breaking,
they nonetheless differ in some fundamental ways from GR.
For example, in GR, the four equations $D_\mu T_{\rm M}^{\mu\nu} = 0$ 
are satisfied by degrees of freedom associated with the matter sector.
However, in theories with explicit diffeomorphism breaking 
$D_\mu T_{\rm LV}^{\mu\nu} = 0$ does not result from the matter dynamics.
It also cannot be imposed by the background fields because they are fixed and
do not allow backreactions. 
Instead, it is the four additional degrees of freedom in the metric that appear
due to the breaking of diffeomorphisms that impose $D_\mu T_{\rm LV}^{\mu\nu} = 0$.

To see this at leading order, consider a paremetrization of the metric as
$g_{\mu\nu} = \tilde g_{\mu\nu} + D_\mu \Xi_\nu + D_\nu \Xi_\mu$,
where $\tilde g_{\mu\nu}$ consists of ten fields that obey four conditions.
Essentially, $\tilde g_{\mu\nu}$ is like a gauge-fixed form of the mertic and $\Xi_\mu$
are the degrees of freedom that would be gauge except for the fact that
the diffeomorphism invariance is explicitly broken.
Inserting this expression in the action,
and using $\tilde g_{\mu\nu}$ in the connection and covariant derivatives,
allows field variations to be performed for the extra degrees of freedom $\Xi_\mu$.
The result is that $D_\mu  T^{\mu\nu}_{\rm LV} = 0$ then holds as the dynamical
equations of motion for the extra metric fields $\Xi_\mu$.

With explicit breaking there are no equivalence classes of 
solutions for the metric as there are in GR.
Instead, definite values of the four additional degrees of freedom $\Xi_\mu$ are
required so as to ensure that $D_\mu  T^{\mu\nu}_{\rm LV} = 0$ holds on shell.
The role of the extra degrees of freedom $\Xi_\mu$ is clearly somewhat unusual.
They do not appear to have any dynamics in their own right.
Instead they acts as buffers between the fixed background $\bar k_\ch$
and the remaining metric and matter fields which can have backreactions.
It is also found that the usual relation between geometry and dynamics
of the matter fields is no longer as clear as it is in GR when there is explicit breaking,
since the extra metric modes $\Xi_\mu$ must play a 
dynamical role as a buffer with the fixed background.

\section{Spontaneous diffeomorphism breaking}

In gravitational effective field theories it is also possible for fixed background tensor
fields to emerge as vacuum expectation values in a process of spontaneous diffeomorphism breaking.
In this case, the background is a vacuum value $\bar k_\ch = \vev{k_\ch}$ of a field $k_\ch$
that is fully dynamical.

At the level of effective field theory,
some models truncate the field $k_\ch$ to its vacuum value $\bar k_\ch$.
However, it is important to keep in mind that $\bar k_\ch$ in this context 
is associated with a dynamical field and that in addition to the vacuum
solutions a complete treatment must also account for the Nambu-Goldstone and massive-mode
excitations about the vacuum solution.\cite{rbak}
In particular,
when these excitations are included, diffeomorphism invariance is recovered in the full action.

It is also important to realize that even when a theory truncates the fields 
$k_\ch$ to their background values $\bar k_\ch$,
these are still vacuum solutions of the equations of motion.
This means that for the case of spontaneous diffeomorphism breaking
the vacuum solution $\bar k_\ch$ obeys
\beq
 \int d^4x 
\sqrt{-g} \fr {\de {\cal L}_{\rm LV}} {\de \bar k_\ch} \de \bar k_\ch
= 0 
\quad\quad ({\rm spontaneous \,\, breaking}) .
\label{sponbreaking}
\eeq
Note that this is in contrast to the result \rf{breaking} for explicit breaking,
where the background does not have to be a solution.
The vanishing of the integral in \rf{sponbreaking} for the case of
sponteaneous breaking eliminates the potential conflict with
the condition in \rf{nobreaking}.
Here, \rf{sponbreaking} holds on shell and there is no conflict with
covariant energy-momentum conservation.

With spontaneous diffeomorphism breaking,
the properties of the resulting theory 
remain similar to those in GR.
The metric still has four gauge degrees of freedom,
and $D_\mu  T^{\mu\nu}_{\rm LV} = 0$ holds as a vacuum solution
for the dynamical field $k_\ch$.
With spontaneous breaking,
the usual linkage between geometry and the dynamics
of the matter fields is mainteined,
and the field $k_\ch$ has backreactions in the form of
Nambu-Goldstone and massive modes.
The main difference between theories with spontaneous diffeomorphism
breaking and GR is that with spontaneous breaking
the vacuum solutions break local Lorentz invariance,
while this does not happen in GR.

An example of an effective field theory that incorporates spontaneous
diffeomorphism breaking is the SME.\cite{sme}
It provides the phenomenological framework for investigations
of Lorentz violation in Minkowski spacetime and in the presence of gravity.
In the gravity sector of the SME, it is assumed that the background
coefficients arise from spontaneous diffeomorphism breaking
and when the Nambu-Goldstone and massive
modes are included the theory is therefore fully dynamical.

\section{Conclusions}

Gravitational effective field theories with explicit 
diffeomorphism breaking are found to be fundamentally different
from GR or theories with spontaneous diffeomorphism breaking. 
With explicit breaking, the fixed backgrounds have no natural physical explanation
and cannot have backreactions.
Instead,
extra modes in the metric must act as a buffer with the
fixed background to ensure covariant energy-momentum conservation.
In contrast, with spontaneous diffeomorphism breaking,
the background fields arise dynamically as vacuum solutions,
and the resulting theories otherwise share many 
of the usual properties of GR.

\end{document}